\documentclass[pre,showpacs,superscriptaddress,twocolumn]{revtex4-1}
\usepackage{amsmath}
\usepackage{amsfonts}
\usepackage{amssymb}
\usepackage{graphicx}
\usepackage{bm}
\usepackage{hyperref}

\usepackage{color}
\usepackage{array,tabularx}
\usepackage{epstopdf}

\def\pf{p_{0}^{\rm pent}}
\def\ps{p_{0}^{\rm hex}}
\def\phl{p_{0}^{\rm hl}}

\begin{document}
\title{The role of cell deformability in the two-dimensional melting of biological tissues}

\author{Yan-Wei Li}
\affiliation{Division of Physics and Applied Physics, School of Physical and
Mathematical Sciences, Nanyang Technological University, Singapore}
\author{Massimo Pica Ciamarra}
\email{massimo@ntu.edu.sg}
\affiliation{Division of Physics and Applied Physics, School of Physical and
Mathematical Sciences, Nanyang Technological University, Singapore}
\affiliation{
CNR--SPIN, Dipartimento di Scienze Fisiche,
Universit\`a di Napoli Federico II, I-80126, Napoli, Italy
}
\date{\today}

\begin{abstract}
The size and the shape of a large variety of polymeric particles, including biological cells, star polymers,
dendrimes and microgels, depend on the applied stresses as the particles are extremely soft.
In high-density suspensions these particles deform as stressed by their neighbours, which implies that the interparticle interaction becomes of many-body type.
Investigating a two dimensional model of cell tissue, where the single particle shear modulus
is related to the cell adhesion strength, here we show that the particle deformability affects
the melting scenario.
On increasing the temperature, stiff particles undergo a first order solid/liquid transition, while
soft ones undergo a continuous solid/hexatic transition followed by a discontinuos hexatic/liquid transition.
At zero temperature the melting transition driven by the decrease of the adhesion strength occurs
through two continuous transitions as in the Kosterlitz, Thouless, Halperin, Nelson, and Young scenario.
Thus, there is a range of adhesion strength values where the hexatic phase is stable at zero temperature,
which suggests that the intermediate phase of the epithelial-to-mesenchymal transition
could be hexatic type.
\end{abstract}

\maketitle

\section{Introduction}

Kosterlitz, Thouless, Halperin, Nelson, and Young (KTHNY)~\cite{Dash,Gasser,KT,HN,Y}
suggested the melting transition in two dimensions to occur through a continuous solid-hexatic transition and
a subsequent continuous hexatic-liquid transition induced by topological
defects: the dissociation of bound dislocation pairs into free dislocations
drives the solid into the hexatic phase, while the unbinding of dislocations
into isolated disclinations drives the hexatic to liquid transition.
Different melting scenarios are however possible~\cite{Chui,Saito,Nelson1978}.
In molecular systems the melting scenario depends on specific system
details, as the softness and the range of pair potential~\cite{Krauth2015,
potential_softness, range_potential}, density~\cite{ningxu}, energy
dissipation~\cite{Tanaka}, shape and symmetry of particles~\cite{Glotzer}, and
so on~\cite{pinned_particle, Qi_pin,Santi, Keim, weikaiqi, Maret1,John_Russo}.
In the prototypical hard disks system melting occurs through a
continuous solid-hexatic transition and a subsequent first-order hexatic-liquid
transition~\cite{Krauth2011,Experiment_harddisc}, while in
systems of hard regular polygons different melting scenarios can be observed
by tuning the number of edges~\cite{Glotzer}.
While the melting of many systems of particles interacting via two-body potentials has been investigated,
nothing is known as concern the features of the two dimensional melting
transition of extended and deformable polymeric particles,
whose shape and volume is not fixed but rather determined by the balance between
the particle mechanical stiffness and the applied stresses.
Particles with these feature are rather common, and include biological cells or
polymeric particles such as star polymers~\cite{Star_polymer, Likos_star},
dendrimers~\cite{Dendrimer}, microgels~\cite{microgels}, polyelectrolyte
stars~\cite{polyelectrolyte_star} and soft granular
particles~\cite{dense_granular, soft_jammed}.

Here we investigate how the deformability of the particles affects the two dimensional
melting transition in the Voronoi model of epithelial cell tissues ~\cite{Frank_vertex,  Staple2010, Bi, Manning1,
Vertex_model}.
Indeed, the melting transition of cell tissues, which occurs in processes involving the movement of
cells such as embryogenesis, tumor spreading, and wound
healing~\cite{Stone2016}, is of particular
interest due to the recent observation
of an intermediate stage in the epithelial (solid-like)-to-mesenchymal
(liquid-like) transition~\cite{EMT1, EMT2, EMT3}.
The Voronoi description of cell tissues is based on an energy functional which has
been derived taking into account the cell
incompressibility and the monolayer's resistance to high fluctuations, as well
as the contractivity of the subcellular cortex and the membrane tension due to
cell-cell adhesion and cortical tension~\cite{Frank_vertex,  Staple2010, Bi,
Manning1,Vertex_model, Bi_prx}.
These two contributions respectively lead to a quadratic dependence of the elastic energy of a cell
$E$ on its area $A$ and on its perimeter $P$, so that
$E = K_A (A-A_{0})^{2}+K_P(P-P_{0})^{2}$.
Here $A_0$ and $P_0$ are preferred values of the cell area and of the cell
perimeter, $K_A$ and $K_P$ are perimeter elastic constants.
In units of $K_A A_0^2$, the energy of the system is
\begin{equation}
e\left(\{\vec r_i\}\right)=\sum_{i=1}^{N}[(a_{i}-1)^{2}+r^{-1}
(p_{i}-p_{0})^{2}],
\label{eq:E}
\end{equation}
where the sum runs over all $N$ cells, $a_i = A_i/A_0$ and $p_i =
P_i/\sqrt{A_0}$. The energy depends on two parameters, the inverse perimeter modulus, $r = K_A
A_0/K_P$, we fix to $r = 1$ if not otherwise stated, and the target shape index,
$p_{0}=P_{0}/\sqrt{A_{0}}$, we will vary.
To investigate Eq.~[\ref{eq:E}] one needs to determine the area and the perimeter
of each cell. In the Voronoi model~\cite{Frank_vertex,  Staple2010, Bi, Manning1,Vertex_model,
Bi_prx} this is done performing a Voronoi tessellation of the
system, and assuming each cell to have the shape of the Voronoi cell associated
to its center of mass. Thus, in this model the centers of mass of the particles are
the degrees of freedom of the system.

Investigating the transition as a function of $p_0$ is of interest
as its increase biologically corresponds to
a decrease of the cell-cell adhesion strength. From a physical viewpoint,
$p_0$ plays two important roles.
On the one side, it fixes the preferred shape of the Voronoi cells.
For instance, for regular hexagons and regular pentagons $\ps \simeq 3.72$ and
$\pf \simeq 3.81$, respectively. More generally, the increase of $p_0$ favours
less compact shapes, and thus a reduction in the number of sides of the Voronoi cells.
On the other hand, $p_0$ also influences the single particle shear modulus,
which in the affine approximation decreases linearly with $p_0$, $\mu = \mu_0
r^{-1} (p_0^*-p_0)$ (see ~\cite{SM}), due to the quadratic dependence of the energy on $p_0$.
$\mu_0$ and $p_0^*$ are constants depending on the shape of the particle.
Thus, larger values of $p_0$ correspond to softer particles.
We remark that this definition of softness concerns the shear modulus of a
single macroscopic particle; it is thus different from the softness investigated
in atomistic models, e.g. Ref.~\cite{Krauth2015}, where softness measures
the contact stiffness of two-body interaction potentials.

Our results show that the
value of $p_0$ affects the melting scenario, as on heating melting occurs via a first-order transition,
at small $p_0$ values, or via a continuous solid-hexatic
and a subsequent discontinuous hexatic-liquid transition,
at larger $p_0$ values. To connect to
the epithelial-to-mesenchymal transition of cell tissues~\cite{EMT1, EMT2, EMT3}, for which thermal
motion is negligible, we have also investigated the $p_0$ dependence
of the zero temperature stable state of the system, and found that
the increase of $p_0$ drives a continuous solid-hexatic transition
and a subsequent continuous hexatic-liquid transition,
as in the KTHNY scenario. This result suggests that the hexatic phase could
be of unexpected biological relevance.

\section{Methods \label{sec:me}}

\subsection{Numerical simulations}
We have performed extensive molecular dynamics simulations of systems of $N$
particles, placed in a rectangular box with aspect ratio $L_{1}:L_{2}=2:\sqrt{3}$ and area $N$, under
periodic boundary conditions.
We employ an in-house massive parallel program to tackle the high computational
cost associated to the need of performing $O(N)$ Voronoi tessellations at each
integration step, to evaluate the forces acting on the particles.
The results presented in the main text, if not otherwise mentioned, refer to
$N=8100$, but we have considered other values of $N$
(\textcolor{black}{up to $16384$ for finite temperature and up to $102400$ for zero temperature}) to show
that finite-size effects are negligible.
We have investigated the melting transition in the NVT ensemble, integrating the
equations of motion via the Verlet algorithm, and fixing the temperature using a
Langevin thermostat~\cite{Allen_book}.
The initial state is prepared relaxing at fixed temperature configurations
prepared by heating/cooling the system.
We check for the convergence of the heating/cooling curves (see Fig.~S6~\cite{SM}), an indication of the
proper thermalization of the system.
Energy minimization have been carried out using
the conjugate-gradient algorithm as implemented in the GNU scientific library~\cite{gsl}.
Voronoi tessellations are computed using the Boost C++ Voronoi library~\cite{boost}.
We remark that the high precision and numerical stability of this library resulted instrumental to correctly
minimize the energy of the system, in particular at high $p_0$ values where the Voronoi tessellations develop degenerate vertexes.

\subsection{Identification of the different phases\label{sec:identification}}
The solid, hexatic and liquid phases have different spatial symmetries one could
reveal investigating the correlation functions of the translational and of the
rotational order parameters.
The positional or translational order parameter is $\psi_T(\vec{r}_j) = e^{i {\vec G} \cdot
\vec{r}_j}$, where $\vec{G}$ is the first peak reciprocal lattice vector of the
triangular crystal, and $\vec{r}_j$ the position of particle $j$.
The rotational or bond-orientational order parameter is
$\psi_{6}(\vec{r}_j)=\frac{1}{n}\sum_{m=1}^{n}\exp(i6\theta_{m}^{j})$, $n$ being
the number of nearest neighbors of particle $j$ and $\theta_{m}^{j}$ being the
angle between $(\vec{r}_{m}-\vec{r}_{j})$ and a fixed arbitrary axis.
In the liquid phase, both the translational and the bond-orientational order are
short-ranged, while in the solid phase both of them are extended, although there
is no long-range translational order in two dimensions~\cite{Mermin}.
In the intermediate hexatic phase, if any, the translational order is
short-ranged, while the bond-orientational one is quasi long-ranged.
\textcolor{black}{The bond-orientational correlation function is calculated as
$g_{6}(r)=\langle\psi_{6}(\vec{r}_{i})\psi_{6}^{*}(\vec{r}_{j})\rangle$ with
$r=|\vec{r}_{i}-\vec{r}_{j}|$. The translational correlation function is calculated as the cut of
the two-dimensional pair correlation function $g(\Delta x, \Delta y)$ along 
the direction $(\Delta x, 0)$ of global orientation of the bond-orientational order parameter~\cite{Krauth2011},
so that the averaging over different configurations is possible.}
Another sensitive characterization~\cite{subblock} of the ordering properties is
provided by the size scaling analysis of the order parameters averaged over
regions of linear size $l_b$,
$\Psi_{T}(l_{b})=\frac{1}{N_{l_{b}}}\sum_{j=1}^{N_{l_{b}}}\psi_{T}(\vec{r}_j)$
and
$\Psi_{6}(l_{b})=\frac{1}{N_{l_{b}}}\sum_{j=1}^{N_{l_{b}}}\psi_{6}(\vec{r}_{j})$, 
where $N_{l_{b}}$ is the number of particles in the considered region.
Indeed, according to KTHNY $\Psi_{T}^{2}(l_{b})/\Psi_{T}^{2}(l) =
(l_{b}/l)^{-x}$ and $\Psi_{6}^{2}(l_{b})/\Psi_{6}^{2}(l)=(l_{b}/l)^{-y}$, with
$x<1/3$ and $y < 1/4$ in the solid phase, $x > 1/3$ and $y < 1/4$ in the hexatic
phase, and $x > 1/3$ and $y > 1/4$ in the liquid phase.
This analysis does not distinguish the coexistence region from the liquid phase,
we discriminate investigating the equation of state.
\section{Results}

\subsection{Finite temperature}

\begin{figure}[tb]
 \centering
 \includegraphics[angle=0,width=0.48\textwidth]{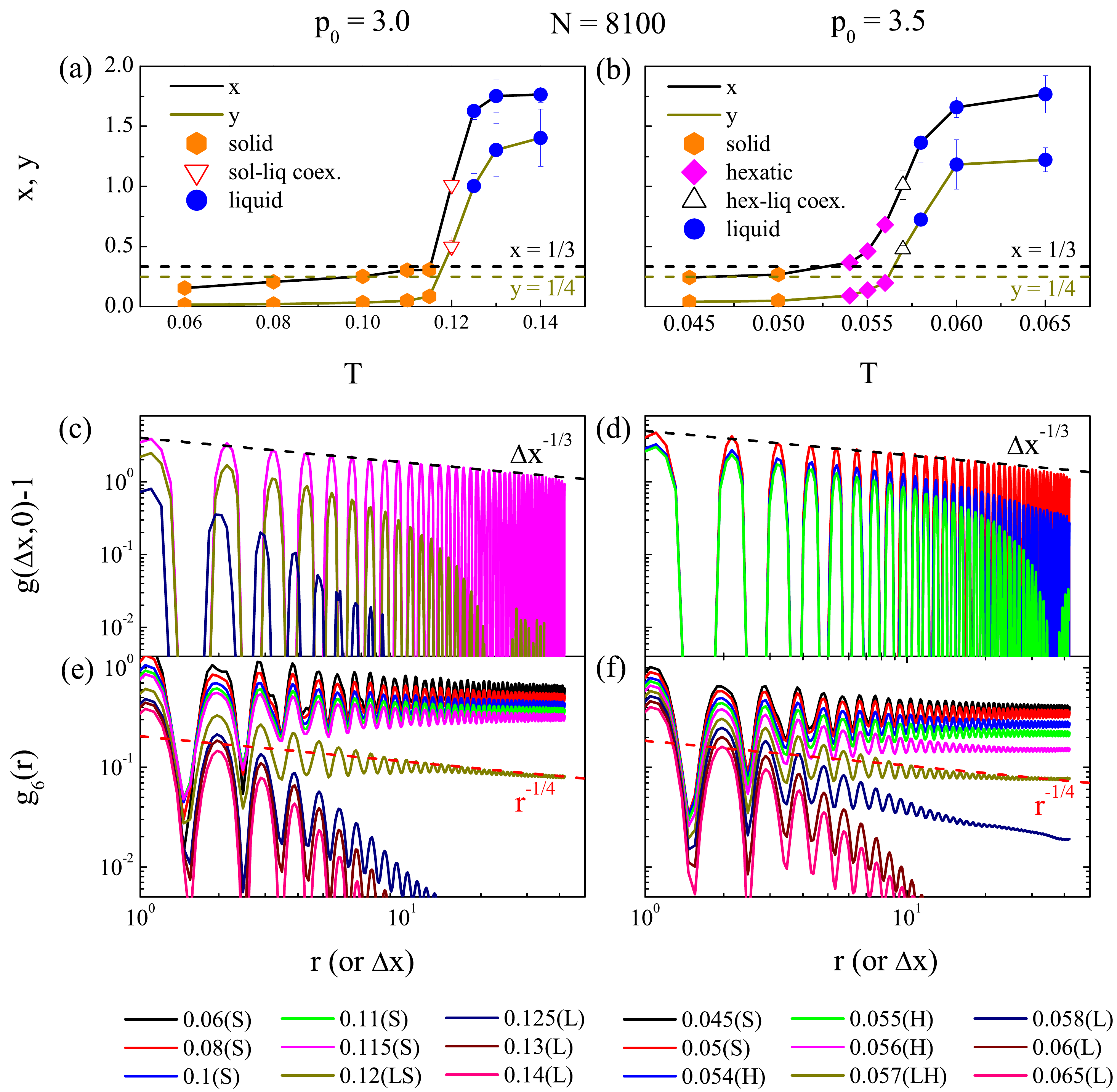}
 \caption{Temperature dependence of the exponents $x$ and $y$ of the sub-block
analysis
  of the translational and bond-orientational order parameters (see text),
  translational correlation function $g(\Delta x, 0)$ and
bond-orientational correlation function $g_6(r)$, for $p_{0}=3.0$
((a), (c) and (e)), and $p_{0}=3.5$ ((b), (d) and (f)).
    The values of the scaling exponents, the decays of $g(\Delta x, 0)$ and of $g_6(r)$, and the
investigation of the equation
  of state (Fig.~\ref{fig:eos}) allow to unambiguously identify the different
phases.
   Depending on the temperature and on $p_0$, states are identified as solid
(S), hexatic (H), liquid-solid coexistence (LS), liquid-hexatic coexistence (LH)
and
 liquid (L), as summarized in the bottom legend.
\label{fig:subblock}
}
\end{figure}

\begin{figure}[!t]
 \centering
 \includegraphics[angle=0,width=0.45\textwidth]{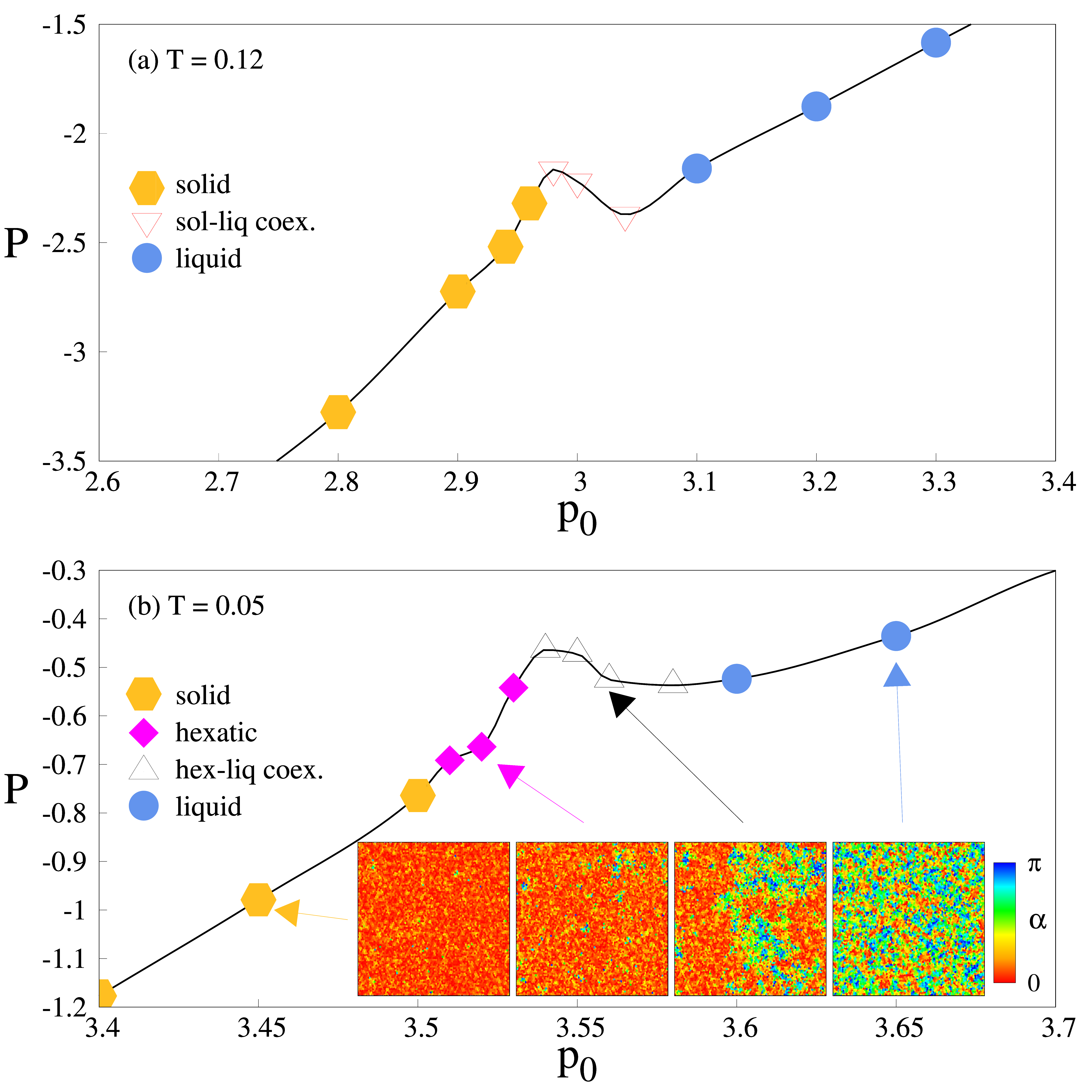}
 \caption{$p_{0}$ dependence of the pressure at (a) $T=0.12$ and (b) $T=0.05$.
Different symbols identify different phases, as in the legends. Lines are to
guide the eye.
For $T =0.05$, snapshots of the system in the different phases are in the insets
of panel (b).
Each cell $i$ is colored~\cite{Krauth2011,ningxu} according to the angle
$\alpha$ between
its bond-orientational order parameter $\psi_{6}(\vec{r}_{i})$ and the global
bond-orientational order parameter
$\Psi_{6}=\frac{1}{N}\sum_{i=1}^{N}\psi_{6}(\vec{r}_{i})$.
\label{fig:eos}}
\end{figure}

Figures~\ref{fig:subblock}(a) and \ref{fig:subblock}(b) illustrate the temperature dependence of the
exponents $x$ and $y$ resulting from the sub-block scaling analysis
(see Sec.~\ref{sec:me} and Fig. S2~\cite{SM}) of the translational and of the bond-orientational order parameters,
for two values of the target shape index. The values of these exponents
allow to identify the different phases, as we discussed in Sec.~\ref{sec:me}.
Specifically, Panel (a) shows that for $p_{0}=3.0$ the melting transition occurs at $T=0.12$
without hexatic phase.
Conversely, panel (b) shows that for $p_{0}=3.5$ there is a temperature range,
$T=0.054-0.056$ of apparent hexatic order.
\textcolor{black}{This result is consistent with the direct investigation of the translational correlation function $g(\Delta x, 0)$ and of the correlation
function of the bond-orientational order parameter $g_{6}(r)$. Indeed, we observe that in the hexatic phase $g(\Delta x, 0)$ decays exponentially, while the bond-orientational order is quasi-long range. In the solid state, $g(\Delta x, 0)$ decays algebraically with an exponent $-1/3$, and $g_{6}(r)$ shows almost no decay, following the KTHNY theory.
The temperature at which $g_{6}(r)\sim r^{-1/4}$ is consistent with the
solid-liquid transition temperature, for $p_{0}=3.0$, and with the
hexatic-liquid transition temperature, for $p_{0}=3.5$.
We have evaluated the translational length scale in hexatic and in liquid phases, and the bond-orientational length scale in the liquid phase, by fitting the corresponding correlation functions with an exponential decay $e^{-r/\xi}$. 
The largest length scale we measured is $\sim20$ (see Fig.~S9~\cite{SM}), a small fraction of the box size ($L_{2}\sim84$), which indicates that the finite size effects may not be strong, possibly because of the soft nature of the system~\cite{Voronoi_liquid}. 
We further support the absence of finite-size effects in Fig.~S7~\cite{SM}, where we compare 
the sub-block plots and the two correlation functions for
$p_0= 3.0$ and $p_0= 3.5$ for the $N = 8100$ system, illustrated in Fig.~\ref{fig:subblock}, with those obtained for
a $N = 16384$ system.}

Since the translational and the bond-orientational order are lost at the same
temperature in the absence of the hexatic phase, and at different temperatures
when this phase is present, we consistently observe in Fig.~S1~\cite{SM} the
associated susceptibilities to peak at the same or at different temperatures.
In Fig.~S5~\cite{SM} we also show that the phase identification from the sub-block analysis
are not affected by the thermally induced shift of the position of the first peak
of the lattice, as this shift is small with respect to the width of the peak (see Fig.~S4~\cite{SM}).
A similar result has been observed in hard polygons~\cite{Glotzer}, while
the shift is relevant in systems of hard disks~\cite{Krauth2011}.

We identify the coexistence region investigating the equation of state,
which we have determined computing the pressure via perturbation techniques~\cite{pressure, pressure1},
due to the many-body nature of the interaction.
Figure~\ref{fig:eos} illustrates the pressure dependence on $p_0$, for two values of the temperature.
We first notice that the pressure is negative, which indicates that the system is under tension.
This is in agreement with our daily experience, as we know that the two margins of a skin cut separate.
However, this is an unusual feature as systems with negative
pressure are generally unstable towards the formation of cavities.
The Voronoi and related models are however
stable at negative pressure as the formation of cavities is hindered as particles
are forced to tessellate the space.
The equation of state shown in Fig.~\ref{fig:eos} illustrates the presence of
Mayer-Wood loops~\cite{Mayer_wood}, which characterize first-order phase
transitions~\cite{Binder}.
Given the phases observed on the two sides of the coexistence curve (see also Fig.~S3~\cite{SM}),
we understand that at $T=0.12$  the liquid and solid phases coexist at $p_{0}\approx3.0$,
while at $T=0.05$ the liquid and the hexatic phase coexist at $p_{0}\approx 3.55$.
On the other hand, there is no pressure loop at the solid-hexatic transition for
$T=0.05$, which is therefore continuous.

\begin{figure}[tb]
 \centering
 \includegraphics[angle=0,width=0.5\textwidth]{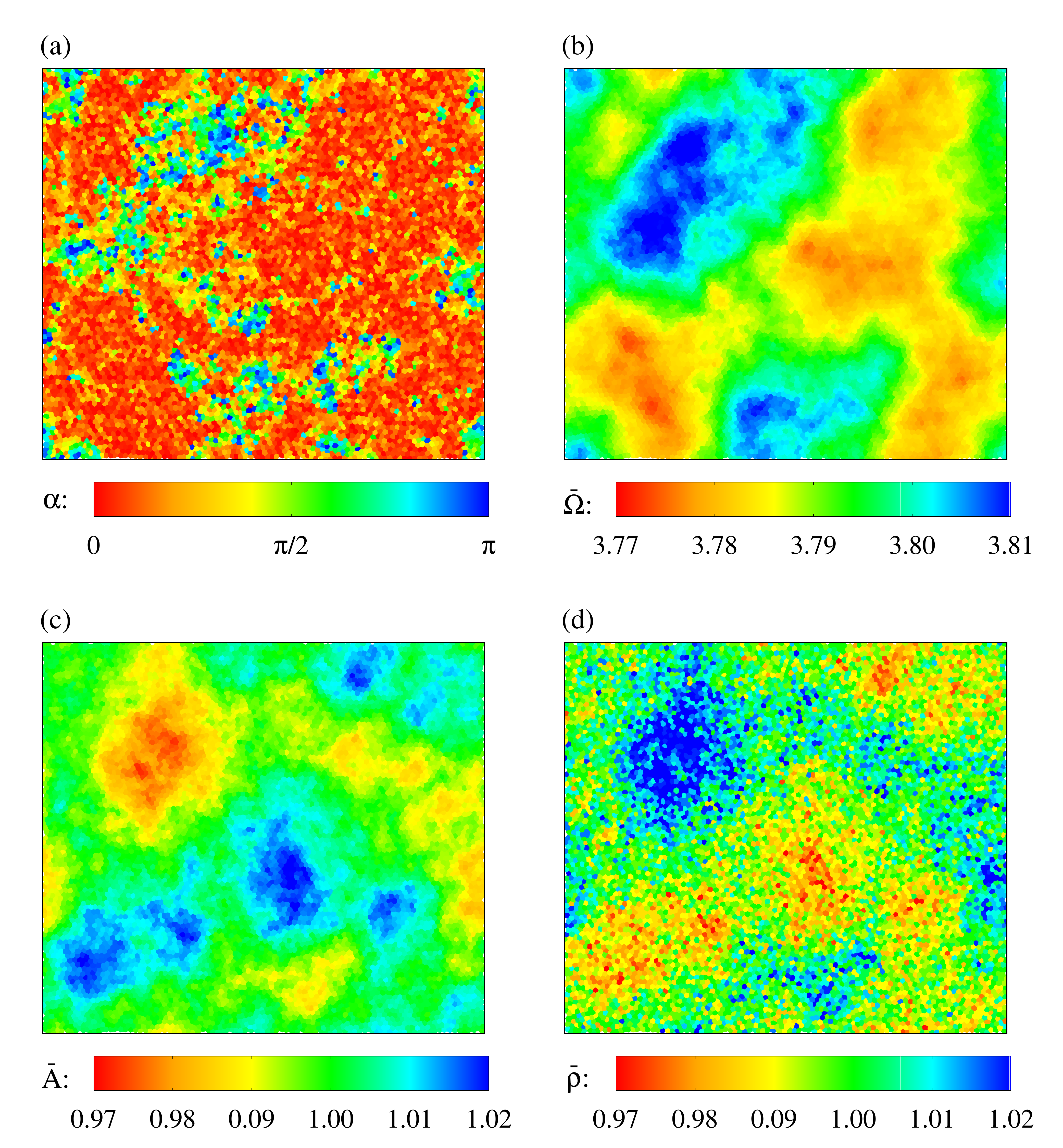}
\caption{Liquid-solid coexistence at $T=0.12$ and $p_{0}=3.0$. The coexistence
is seen by coloring the particles according to (a) the angle $\alpha$ between
$\psi_{6}(\vec{r}_{i})$ and $\Psi_{6}$, (b) the local perimeter-to-area ratio,
(c) the local area, and (d) the local number density.
\label{fig:coex}}
\end{figure}

As an additional check of the identified phases we have also visualized the system
by coloring each cell according to its properties.
For instance, the bond-orientational ordering is visualized by
associating to each particle $i$ a color related to the angle $\alpha$ between
the local, $\psi_{6}(\vec{r}_{i})$, and the global, $\Psi_{6}$, order
parameters. Examples are in the insets of  Fig.~\ref{fig:eos}(b), which shows that the 
identified phases are consistent with their real space observation.
The phase coexistence is also visualized by investigating the spatial distribution
of other quantities, such as the average local perimeter-to-area ratio
$\overline{\Omega}$, the average local area $\overline{A}$, and the average
local number density $\overline{\rho}$.
Here, we define these quantities by averaging over a circular region of radius
$r_c = 10$~\cite{Glotzer}.
Interestingly, cells exhibiting solid-like behavior (red-colored
regions in Fig.~\ref{fig:coex}(a)) have low perimeter-to-area ratio (Fig.~\ref{fig:coex}(b)), high local
area (Fig.~\ref{fig:coex}(c)), and low number density (Fig.~\ref{fig:coex}(d)). That is, the solid is less dense than the liquid.
This unusual feature also occurs in liquids with density anomalies,
such as water, and in systems with a re-entrant melting transition~\cite{Likos-reentrant}.

\begin{figure}[!!t]
 \centering
 \includegraphics[angle=0,width=0.49\textwidth]{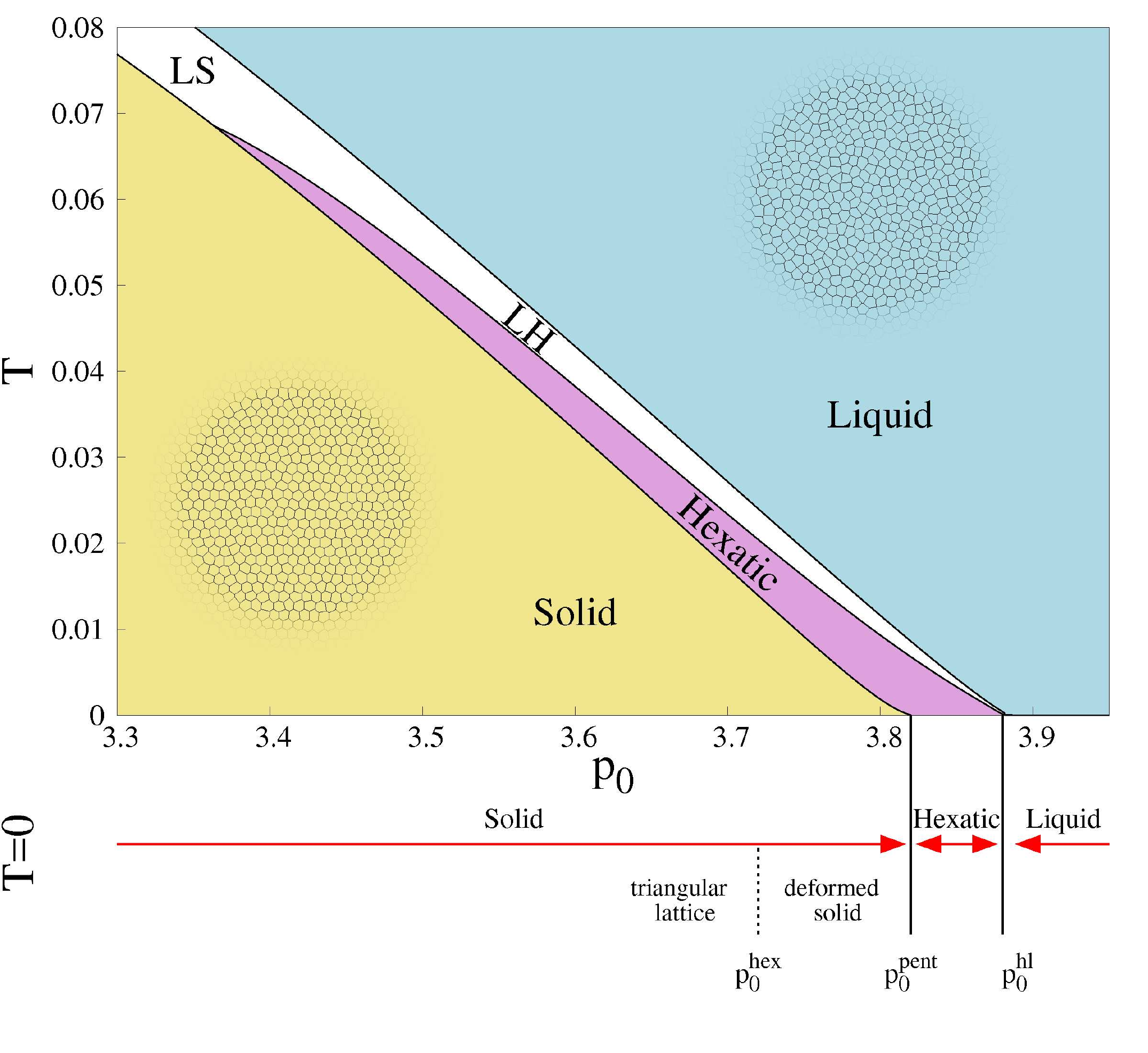}
 \caption{Phase diagram as a function of
temperature $T$ and target shape index $p_{0}$.
The lines interpolate the estimated boundaries between the different phases.
The lowest temperature at which we are able to equilibrate the system
depends on $p_0$, and varies from $T \simeq 10^{-2}$ at small $p_0$ values,
to $T \simeq 10^{-4}$ at large $p_0$ values.
The bottom part of the figure illustrates the properties of the zero temperature stable state
as a function of $p_0$. The insets illustrate typical configurations in the two
phases.
\label{eq:pd}}
\end{figure}

The upper part of Fig.~\ref{eq:pd} illustrates the melting phase diagram resulting
from this investigation.
At small $p_0$ values the system melts through a first order transition,
while a two-step melting with a continuous solid-hexatic and a consecutive first-order hexatic-liquid
transition appears for $p_{0} \gtrsim 3.4$.
While the identification of this crossover value of $p_0$ is difficult, we have carefully verified
for the absence of the hexatic phase at small $p_0$ values by checking the phases
observed at the boundaries of the coexistence curve, we determine through a Maxwell construction
(see Fig.~S8).
The absence of a hexatic phase is also
consistent with the fact that susceptibilities associated to the positional order and to the
bond-orientational peak at the same temperature at small values of $p_0$
(see Figs.~S1(b), S1(d) and S1(f)).
\textcolor{black}{The lowest temperature at which we were able to equilibrate the system in the hexatic 
phase is $T=0.0002$, at $p_{0}=3.84$. Extrapolating the solid/hexatic transition line to zero temperature we find
the two-step melting scenario to extend up to $p_0 \simeq \pf$. 
}
For $\pf\leq p_{0}<\phl$ the zero temperature
state appears to be of hexatic type, and we therefore only observe a first-order hexatic-liquid transition.
Due to the narrowing of the coexistence region, we do not exclude the presence
of a continuous hexatic-liquid transition at $p_{0}$ values very close to $\phl$.

\subsection{Zero temperature}
\begin{figure*}[tb]
 \centering
 \includegraphics[angle=0,width=0.7\textwidth]{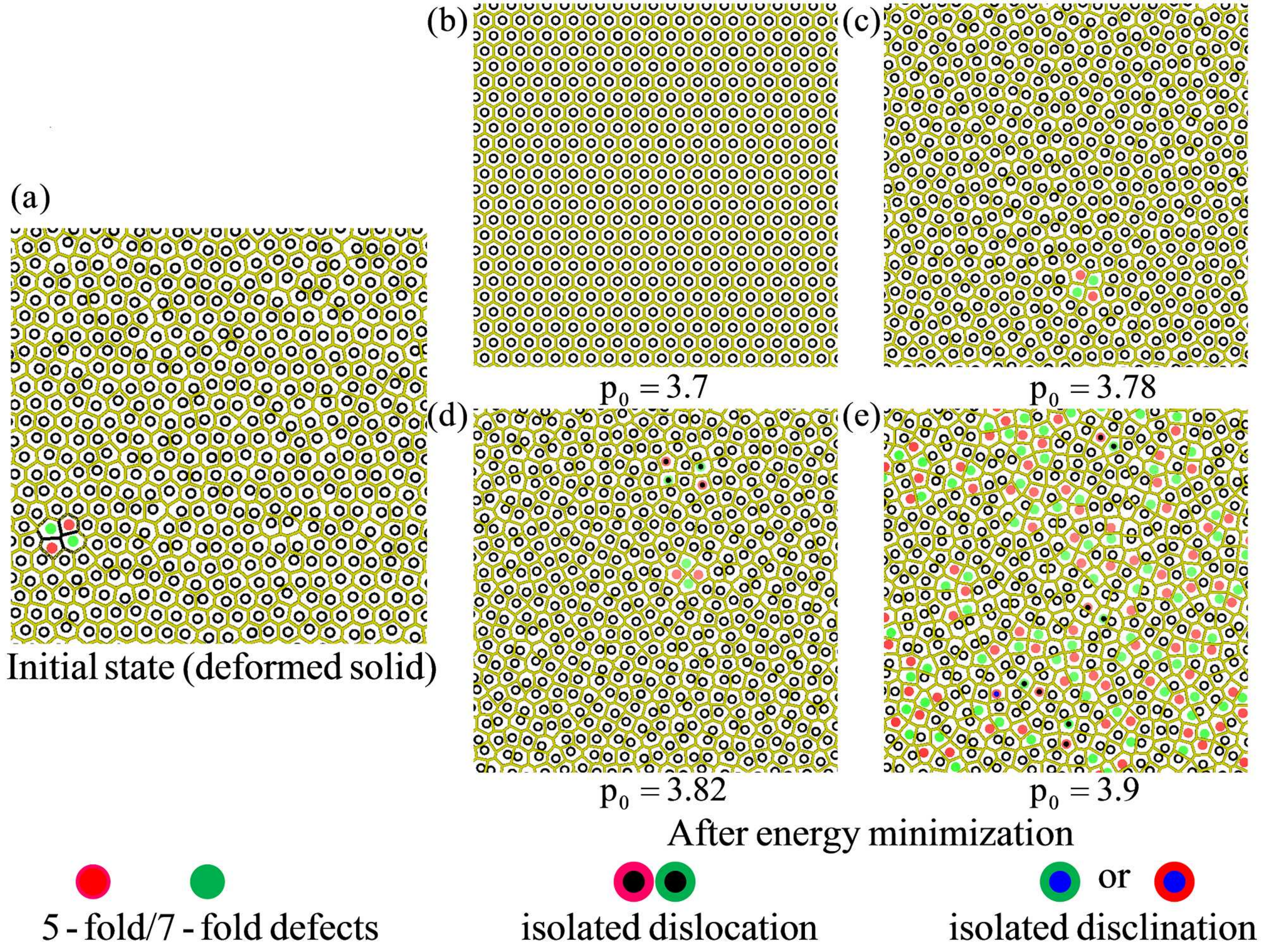}
\caption{The energy minimization process of a configuration
in the deformed solid state (a) can end-up in configurations
belonging to different phases, depending on the value of $p_0$.
The hollow circles represent particles with six nearest neighbors, i.e., the
hexagon-shaped cells. Red filled circles are 5-fold defects and green filled
circles are 7-fold defects. Isolated dislocations are marked by smaller black
dots and isolated disclinations are illustrated with smaller blue dots.
The snapshots illustrate only a small part of our investigated system.
\label{fig:min}}
\end{figure*}

The phase diagram of Fig.~\ref{eq:pd} suggests the existence of a range of $p_0$ values
where the hexatic state is the stable state at zero temperature. This is an interesting result, as so far
the hexatic phase has only been observed at finite temperature. In addition,
this suggests that cell tissues, that are not thermal systems, might actually have a stable hexatic phase.
We further investigate this matter evaluating the properties of
zero-temperature configurations generated by minimizing the energy of the system using the conjugate
gradient protocol (see Sec.~\ref{sec:me}). \textcolor{black}{As in the thermal case, we identify
the phase of these configurations performing the sub-block scaling analysis (see Sec.~\ref{sec:me}) 
of the order parameters.} 
We have considered minimization starting from different
initial conditions: perfect hexagonal lattice, deformed solid, hexatic, liquid, random configuration,
and square lattice. Here by deformed solid we indicate a configuration
with the symmetries of the hexagonal lattice, in which the cells are not perfect hexagons,
alike a finite temperature hexagonal lattice configuration (see Fig.~\ref{fig:min}(a)).
The minimization terminates when the average energy per cell changes by less than $10^{-7}$ between consecutive
minimization steps, which is our criterion for convergence, or after $10^{4}$ steps.
When the initial state is of deformed solid type, the minimization algorithm
converges if $p_0$ is not in the $[\ps, \pf]$ range corresponding to the deformed solid
configuration. We remark, however, that even when convergence is not achieved
the geometrical features of the system appear to be clear as
particles move by minute distances.
However, this may not be the case for the mechanical properties of the model~\cite{Daniel_2017}.

\begin{figure}[tb]
 \centering
 \includegraphics[angle=0,width=0.4\textwidth]{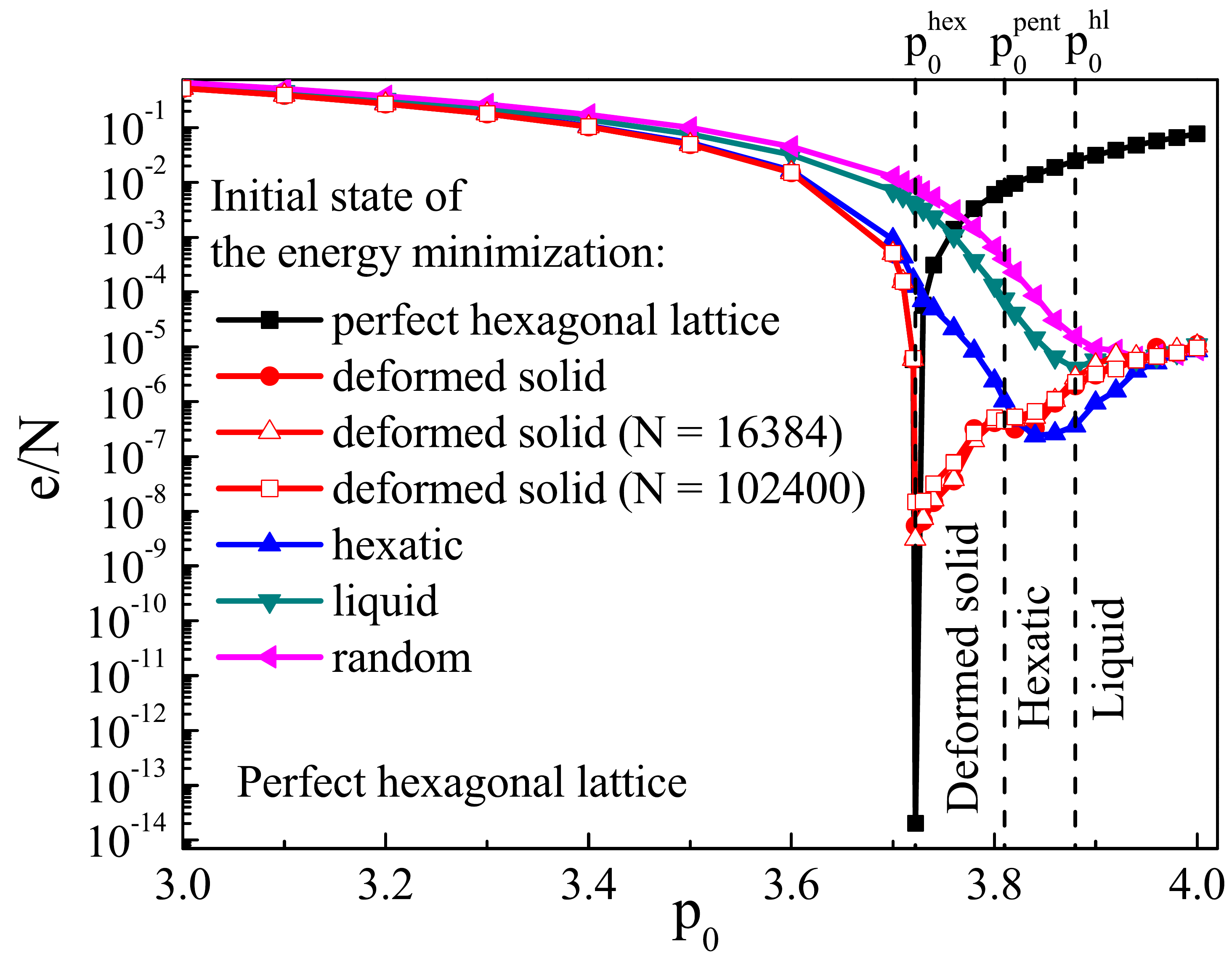}
\caption{$p_{0}$ dependence of the average energy per cell after performing
the energy minimization algorithm. Different curves refer to minimization carried
out starting from initial states with different symmetries and system sizes, as in the legend.
\label{fig:eminima}}
\end{figure}

We show in Fig.~\ref{fig:eminima} the $p_{0}$ dependence of the average energy
per cell of the final state reached by our minimization procedure.
The energy curve obtained minimizing the energy of the perfect hexagonal lattice
has a non-monotonic dependence, as it vanishes at $p_{0}=p_{0}^{\rm hex}$.
For smaller values of $p_0$, the zero temperature stable state is the hexagonal lattice,
which is the final state reached when minimizing the energy starting from hexagonal, deformed
solid or hexatic configurations (e.g., see Fig.~\ref{fig:min}).
We also observe that the deformed solid has the lowest energy in the $p_{0}$
range $p_{0}^{\rm hex}<p_{0}<\pf$, while the hexatic has the lowest energy in the
$p_{0}$ range $\pf<p_{0}<\phl\simeq 3.88$.
Both the deformed solid and hexatic configurations transform into liquid-like configurations
if the energy is minimized with $p_{0}\ge\phl$.
For this large value of $p_0$, all of the energy curves converge except for the one
obtained from the perfect hexagonal lattice, clearly indicating that the stable state
is in the liquid phase and that the hexagonal phase is metastable.
We notice that in the vertex model, where the tesselation is not forced to be of Voronoi type,
the transition from a perfect hexagonal lattice to a
disordered soft lattice at $p_{0}=\ps$ has also been previously rationalized~\cite{Staple2010}.
We also remark that at $p_0 = 4$ the energy of the system vanishes for a square lattice configuration.
We have checked that energy minimizations starting from perfect or slightly
distorted square lattices do not lead to states with energy smaller than the one
we obtain minimizing the energy starting from a deformed solid configuration,
unless $p_0$ is very close to $4$ (approximately, $|p_0-4| < 0.05$). Thus,
there is a small $p_0$ interval around $p_0=4$ where the zero temperature stable phase has the square symmetry.

\begin{figure}[tb]
 \centering
 \includegraphics[angle=0,width=0.4\textwidth]{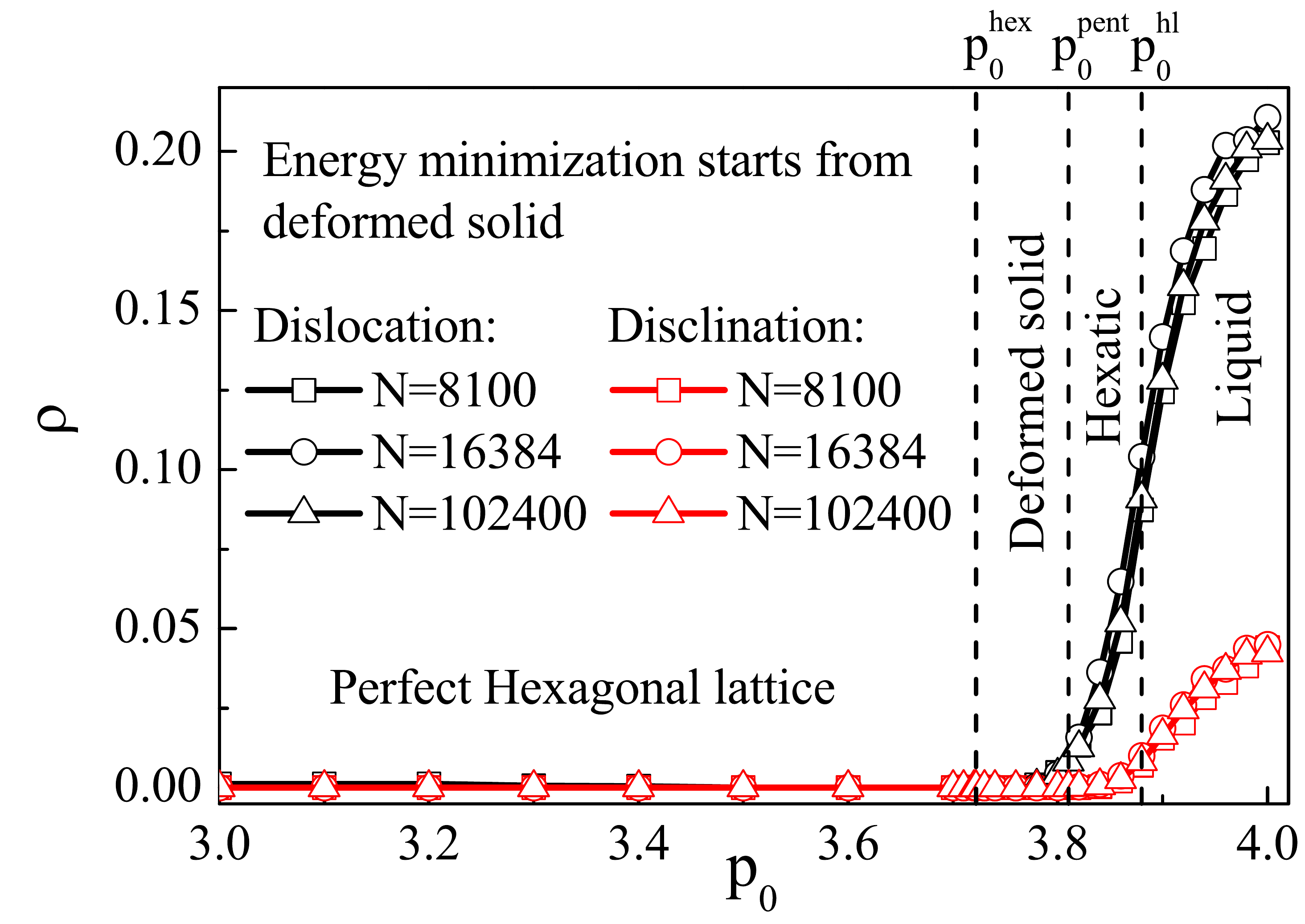}
\caption{$p_{0}$ dependence of the fraction of dislocations and disclinations of
the zero-temperature states of the model with the smallest energy.\label{fig:rhodefects}}
\end{figure}

The vertical dashed lines in Fig.~\ref{fig:eminima} summarize the emerging zero-temperature phase diagram.
We also illustrate this zero-temperature diagram in the lower part of Fig.~\ref{eq:pd},
to stress that this is consistent with the zero temperature extrapolation of the finite temperature phase diagram.
While the location of the different phase boundaries might be protocol dependent, the results
appear to be robust as the system is able to change phase during the minimization.
Specifically, we observe the following transitions in the relevant $p_0$ ranges during the minimization process:
hexatic $\to$ deformed solid; deformed solid $\to$ hexatic; deformed solid and hexatic $\to$ liquid;
deformed solid and hexatic $\to$ perfect hexagonal lattice (see Fig.~\ref{fig:min}).
In Fig.~\ref{fig:eminima} and Fig.~\ref{fig:rhodefects} we also
reports results obtained for a larger system size, to stress that these zero temperature
results do not suffer from finite-size effects.

\textcolor{black}{At zero temperature, we have found no evidence of phase coexistence, 
indicating that the melting transition driven by an increase of $p_0$ occurs through two
consecutive continuous transitions, as in the prototypical KTHNY scenario.
According to this scenario, the two transitions are mediated by topological defects,
the solid-hexatic transition corresponding to the dissociation of bound dislocation
pairs into free dislocations, the hexatic-liquid corresponding to the
unbinding of dislocations into isolated disclinations.
We confirm this topological interpretation in Fig.~\ref{fig:min}
which illustrates the defects characterizing the different phases.
In the perfect hexagonal lattice, there are no defects.
In the deformed solid phase, defects are mainly bounded dislocation pairs
(5-7-5-7 quartets).
In the hexatic phase, there are isolated dislocation (5-7 pairs) and no isolated
disinclination (5-fold or
7-fold defects).
Finally, in the liquid phases defects are both isolated disclinations and
isolated dislocations.}

\textcolor{black}{We have validated this topological interpretation
investigating the $p_0$ dependence of the fraction $\rho$
of dislocations (both isolated and clustered) and of the fraction of
disclinations of the local energy minima
generated from deformed solid configurations.
Figure~\ref{fig:rhodefects} illustrates the results, that do not depend on the system size.
As expected, we find no defects in the perfect hexagonal lattice.
The fraction of cells in dislocations starts
increasing at $p_{0}\simeq\pf$, above which the zero temperature phase changes
from deformed solid to hexatic. For $\pf<p_{0}<\phl$,
where the zero temperature phase is hexatic, the fraction of
particles in dislocations sensibly grows, while the fraction of particles in
disclinations remains low. At $p_{0}>\phl$, where the system enters the liquid phase,
both the fraction of cells in dislocations and that in disclinations grow.}

We have also investigated the zero-temperature phase diagram for a different
value of the inverse perimeter modulus, $r=100$,
and found it at most weakly dependent on $r$.
An analogous result has been reported in the investigation of the vertex
model~\cite{Bi}, where $\pf$ marks the location of a rigidity transition regardless of the value
of $r$.
We remark that the most interesting aspect of the zero-temperature phase diagram is the
presence of a range of $p_0$ values in which the minimal energy state is the hexatic one, a feature that to
our knowledge has never been observed before. This hexatic phase could be possibly associated
to the intermediate stage of the epithelial to mesenchymal transition, which occur at similar values of
$p_0$~\cite{Park}.

\section{Discussion}
We rationalize the temperature dependence of the transition lines of the thermal
phase diagram considering that $p_0$
affects the shear modulus of the particles, and hence that of the system.
Indeed, as the single particle shear modulus (see ~\cite{SM}),
also the macroscopic modulus $\mu$ decreases linearly as $p_0$ increase, to a good approximation~\cite{Bi}.
Considering that according to the Lindemann criterion the melting transition occurs at a
temperature $T_m \simeq \mu\lambda^2$, with
$\lambda$ a microscopic distance, e.g. a fraction
of the interparticle separation, this explains the approximately linear dependence of the transition temperatures on
$p_0$.
We also note that, since in the affine approximation the single particle shear
modulus is proportional to $r^{-1}$,
the transition temperatures are expected to also scale as $r^{-1}$.
Whereas the melted solid is in the hexatic or the liquid phase depends on the
free energy difference between these two phases,
which is $\Delta f(p_0,T) = u_{\rm dis}(p_0)-Ts_{\rm dis}$,
where $u_{\rm dis}(p_0)$ is the energy required to unbind the dislocations,
and $s_{\rm dis}$ the corresponding entropy gain.
$\Delta f(p_0,T_h) = 0$ identifies the limit of stability of the hexatic phase,
that occurs as long as $u_{\rm dis}(p_0)>Ts_{\rm dis}$.
As $p_0$ increases the hexatic phase becomes unstable as the unbinding energy
decreases. In particular, since the range of $p_0$ values
in which the hexatic phase is stable at zero temperature extends up to $p_0 = \phl$,
we understand that $u_{\rm dis} < 0$ for $p_0 > \phl$.
The entropic contribution to $\Delta f(p_0,T)$ drives the hexatic to liquid
transition on increasing the temperature.

\begin{figure}[!tb]
 \centering
 \includegraphics[angle=0,width=0.4\textwidth]{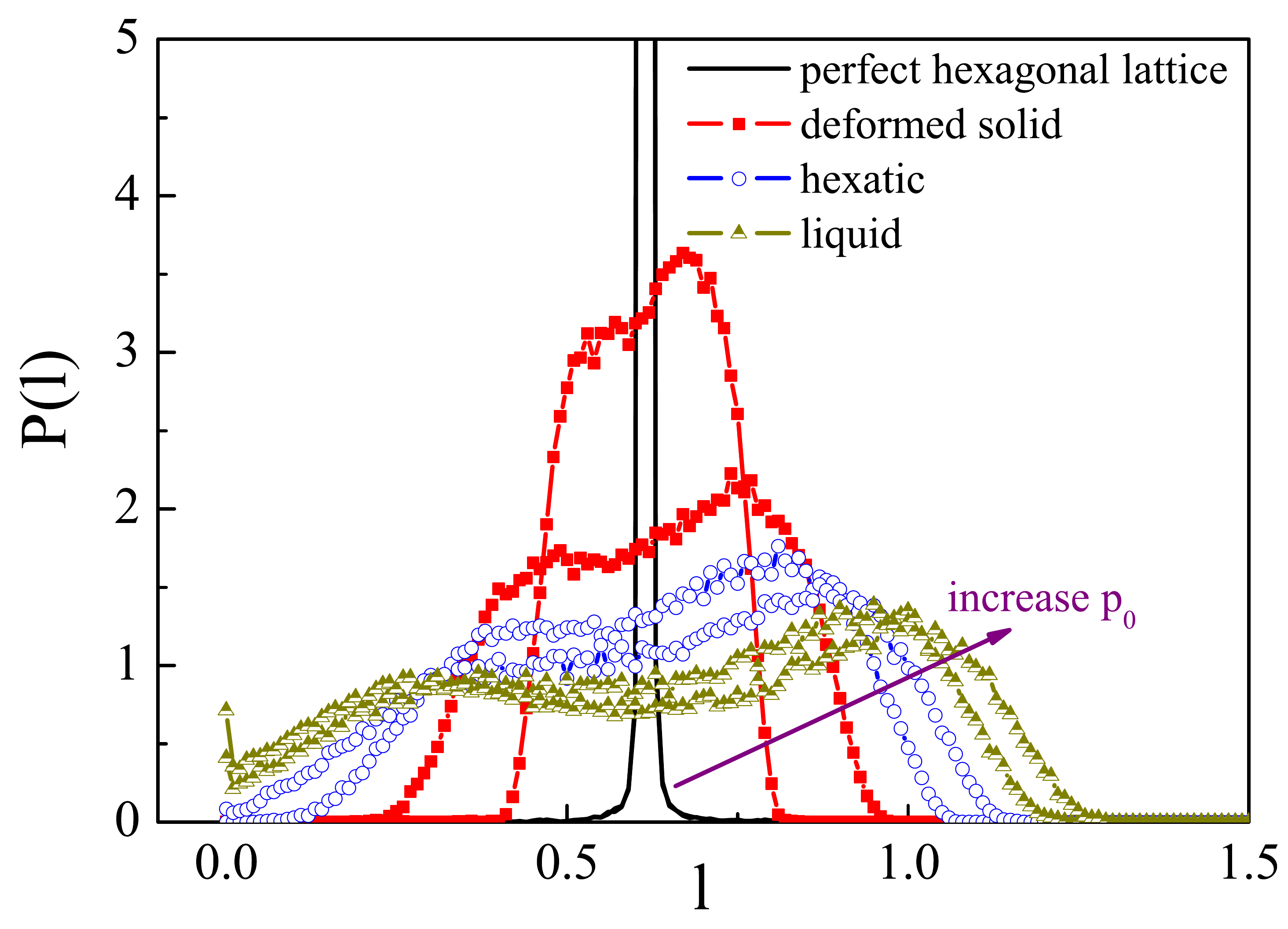}
\caption{$p_{0}$ dependence of the probability distribution of the lengths of the Voronoi edges
of the zero-temperature states of the model with the smallest energy.
\label{fig:PL}}
\end{figure}

A qualitative understanding of the features of the zero temperature phase diagram
is obtained considering that $p_0$ controls the degree of frustration of the system.
Specifically, given that for the hexagonal crystal $p_0 = \ps$,
$|p_0-\ps|$ qualitatively acts as a degree of frustration of the hexagonal crystalline state,
as also clear from Fig.~\ref{fig:eminima}.
The increase of the degree of frustration of a system generally makes
disordered the stable state.
Accordingly, it is surprising
that for $p_0 < \ps$ the hexagonal lattice remains the zero temperature phase, as if frustration was absent.
Conversely, for $p_0 > \ps$ different phases are observed, as expected.
We rationalize these observations considering that for $p_0 < \ps$,
the hexagonal crystal is frustrated as the particles should be more compact than the hexagons,
a feature generally associated to polygons with many sides.
However, since the system tessellates the space,
states with an average number of sides per cell larger than 6 cannot
exist because of Euler's theorem for planar graphs. We therefore understand that the model has a topological
constraint that effectively prevents frustration to play its usual role for  $p_0 < \ps$.
The same argument does not apply for $p_0 > \ps$, where cells should be less compact than
the hexagonal ones, and thus with an average number of sides smaller than $6$.
Thus, for $p_0 > \ps$ frustration plays its usual role.
Indeed, the deformed solid region can be seen as a region of small frustration,
which is not able to sensibly affects the zero temperature phase.
Further increases of $p_0$, conversely, makes the zero temperature phase first of hexatic and
then of liquid type.
This scenario is consistent with the features of the distribution the lengths of the Voronoi edges, $P(l)$,
illustrated in Fig.~\ref{fig:PL}.
In the crystalline phase, as expected, $P(l) = \delta(l-\ps/6)$, regardless of the value of $p_0$.
The distribution broadens and develops a bimodal shape in the deformed solid phase and in the hexatic one.
Finally, in the liquid phase $P(l)$ develops a delta-peak in zero, consistently with the emergence of a growing number of degenerate vertices.

Recent experimental studies have shown the existence of an intermediate phase in the
epithelial to mesenchymal transition~\cite{EMT1,EMT2, EMT3}, which occurs
at measured value of $p_0$ consistent with those found in the Voronoi model~\cite{Park}.
This suggests that the observed intermediate phase could be of hexatic type. In this respect, it also worth noticing that a moderate polydispersity,
which is expected to occur in actual tissues, does not affect the melting scenario~\cite{Frenkel}.
We hope our results will stimulate relevant experimental work to test the nature of the intermediate
epithelial-mesenchymal phase.
Our results also suggests the possibility of using a Voronoi model with a different energy functional
to investigate different systems of soft deformable particles at high density.
Soft polymeric particles such as microgels, for instance, shrink and deform when compressed~\cite{Cloitre,Romeo},
which might help their crystallization~\cite{Scotti2016}.
In particular, neutral or large enough particles, that do not undergo a
counter-ion induced deswelling~\cite{Scotti2016,Kobayashi2016},
might tessellate the space at large enough densities.
This line of research might offer a novel approach
to rationalize some unusual features of these systems,
such as their high diffusivity~\cite{Mattsson2009} in a region of very high density,
where particles are highly deformed.
Finally, we wish to remark that this model is the first one that has been shown to exhibit a hexatic phase
at zero temperature. Possibly, the investigation of the zero temperature
melting transition could help rationalizing previous results that have
found many parameters to affect the melting
scenario in different systems~\cite{Krauth2011,Glotzer,Tanaka,John_Russo,pinned_particle,Qi_pin,
Krauth2015,ningxu,Santi,potential_softness,range_potential}.

\begin{acknowledgments}
We acknowledge support from the Singapore Ministry of Education through the
Academic Research Fund (Tier 2) {\it Many-Body Interactions In Concentrated Colloidal Suspensions}
MOE2017-T2-1-066 (S), and are grateful to the National Supercomputing Centre (NSCC) of Singapore for
providing computational resources.
We thank R. Ni, D. Bi and Y. Zheng for helpful discussions.
\end{acknowledgments}


%

\end{document}